\documentclass[12pt]{article}
\begin{document}
\title{An Abstract Interface to Higher Spin Gauge Field Theory}         
\author{A. K. H. Bengtsson\footnote{e-mail: anders.bengtsson@hb.se. Work supported by the Knowledge Foundation.}}

\maketitle
\begin{center}School of Engineering, University College of Bor\aa s,\break All\' egatan 1, S-50190 Bor\aa s, Sweden.\end{center}

\begin{abstract}
A comprehensive approach to the theory of higher spin gauge fields is proposed. By explicitly separating out details of implementation from general principles, it becomes possible to focus on the bare minimum of requirements that such a theory must satisfy. The abstraction is based on a survey of the progress that has been achieved since relativistic wave equations for higher spin fields were first considered in the nineteen thirties. As a byproduct, a formalism is obtained that is abstract enough to describe a wide class of classical field theories. The formalism, viewed as syntax, can then be semantically mapped to a category of homotopy Lie algebras, thus showing that the theory in some sense exists, at least as an abstract mathematical structure. Still, a concrete physics-like, implementation remains to be constructed. Lacking deep physical insight into the problem, an implementation in terms of generalized vertex operators is set up within which a brute force iterative determination of the first few orders in the interaction can be attempted.
\end{abstract}
\pagebreak

\section{Introduction}\label{sec: Introduction}
On a high enough level of abstraction, the theory of self-interacting higher spin gauge fields become either trivial or void. This might seem like a preposterous statement about a problem, the solution of which has eluded theoretical physics since the nineteen seventies, when the problem was first explicitly raised \cite{FangFronsdal1979}. In this paper, I will try to explain the intuition behind this claim. 

Theoretical physics in general, and high energy physics in particular, rest on, as is well known, a tremendous body of knowledge about reality. The nature of this knowledge is manifold, one aspect of it is the general principles like the relativity principle, the equivalence principle, various gauge principles and of course the quantum paradigm. Another aspect is the many detailed and elaborate calculational schemes employed in particular models, realistic or of the 'toy' variant. This is the 'nuts and bolts' of the science. Whereas the principles are lofty and beautiful to contemplate, the nuts and bolts are often ugly and boring to struggle with. Of course, this is a matter of taste and outlook. But in the end, the nuts and bolts must be there in the right place in order to make contact with experiment, and ensure eventual mathematical consistency. 

In computer science, we also find this division between high level abstract approach to problems, and low level nuts and bolts code grinding. But in computer science the division is more explicitly pronounced. The complexity of modern software development has forced an approach where one has to get the principles right first. 

One purpose of the present paper is to adopt this mode of working with respect to the problem of introducing self-interactions among higher spin massless gauge fields. Substantial progress notwithstanding, the problem is still not completely solved, and far less understood. It is not even clear how to recognize or evaluate a purported solution. Massless higher spin fields appear in many contexts related to string theory, membrane theory and M-theory and theories deriving from these. This makes it interesting to find out whether higher spin gauge fields can stand on their own, without crutches, so to speak, from circumstantial theoretical constructs. Furthermore, a problem so simple to formulate, but so difficult to solve, is intriguing in itself.

I will approach the problem by formulating an as general as possible {\it interface} to higher spin gauge field theory. In the process, specifications that the theory has to meet, will be cataloged. The interface will then turn out to be quite trivial. Then comes the question of actually implementing the interface. This is where the nuts and bolts enters. Perhaps there is no implementation, then the theory is void.

Now, what do we gain by adopting this strategy? Firstly, we get a framework where we can discuss the general overall aspects of the theory without worrying at the same time whether they are implementable or not. We don't have to fix space-time dimension or signature, or worry about background geometry and coupling to gravity. Indeed, we don't have to worry about spacetime at all. Secondly, it might be possible to separate the issue of existence from the issue of construction. Thirdly, if the theory exists, there might be several, in some sense, different implementations, thus avoiding the so common pattern of thinking in terms of uniqueness. Fourthly, it might be possible to actually implement the theory computationally in some set of abstract data types.\footnote{In that case, considerations of finite definition enters, but that can presumably be taken care of relying on lazy evaluation, thus effectively allowing denumerably infinite data structures.} And lastly, and perhaps not completely independent of the previous points, we can separate physics from mathematics. The problem is so difficult that we cannot a priori know which of our cherished physical principles that can be retained. Better then to keep an open mind and treat the problem purely mathematically.

There is one further point to be made. When solving hard theoretical physics problems it is natural to search through mathematics in the hope of finding a pre-existing structure that can be deployed. However, one could imagine a physics problem for which there is no mathematical structure available as yet. That this situation could be encountered in fundamental physics is not at all unthinkable. It seems to me that computer science (CS) has tools to tackle this situation, or at least, formulate it. Thus, again borrowing from CS, the approach to higher spin theory proposed here can also be viewed as an attempt to provide a {\it syntax} for the problem. Then implementation would correspond to providing the syntactical model with {\it semantics}. Now syntax and semantics are concepts normally applied to programming languages, or formal languages in general, so I'm using the concepts in a slightly transferred sense. Continuing this train of thought, of the different semantical schemes, {\it denotational} semantics, where the syntactical structures are mapped to mathematical objects in a pre-existing (and well understood) semantical domain {\cite{Stoy}, seems to be the most appropriate. If a semantical domain cannot be found, then research should perhaps be directed towards inventing new mathematics, rather than trying to solve a theoretical physics problem. If the domain exists, we might be able to target this mathematical structure more precisely. This is one reason that I'm adopting CS inspired thinking rather then working directly within a mathematical structure from the outset. If it seems strange to use computer science concepts in high energy physics, one should consider the circumstance that the subject of theoretical computer science is really data and processes in general, and therefore it can be useful in various scientific contexts where one has to deal with complex systems.

Parts of the discussion in this paper is quite elementary. That is inherent in the formal, syntactic approach. I want to focus on the abstract and general issues involved, not taking, at least not consciously, to much pre-existing mathematics on board. I find it unlikely that the higher spin problem can be solved by unguided index hacking, no matter how stubborn. But as the present work is mainly conceptual and in a creative phase, the formalization will not be pushed to far. A pure syntactic approach would almost certainly obscure the main idea. In order to communicate, I will compromise by using a somewhat unprincipled mix of syntax, semantics and mathematics. If the approach is fruitful, an exact formulation can always be set up later.

Perhaps a simple example helps to further explicate my point of view. Consider the real numbers. The axioms for this structure are well known and used almost subconsciously in everyday calculating. When working abstractly in, say, calculus or real analysis, solving differential equations or whatever, these axioms and the theorems are used throughout. We never worry about their relevance or their truth. But of course, if there weren't any implementation of the axioms, the exercise would be void, i.e. just formal manipulations. Now we know that there are different implementations of the real numbers in terms of for example Dedekind cuts or limits of Cauchy sequences. These implementations are in their turn based on implementations of the rational numbers in terms of the natural numbers. The story is well known. But one more point can be made. If numerical calculations have to be done, then a detailed implementation of the real numbers in terms of floating point numbers is necessary.

Now, of course, we never start with a complete blank mind. We have some knowledge about the problem at hand and we often have more or less strong intuitions. There might also be folklore on the subject, but in my opinion, folklore is often too prejudiced to be useful in a creative way.

So the first step will be to abstract from what we already know. The scope of the present paper is therefore to set up a general enough framework within which the problem of self-interactions can be analyzed, while allowing for various forms for the free theory and allowing for different implementational ideas as regards interactions.

Two other sources of ideas for the present paper should be mentioned. One is string field theory, \cite{Witten1986a,CBThorn1989}, in particular the non-polynomial closed string field theory \cite{KugoSuehiro1990,Zwiebach1993a}, and the other is the mathematical theory of higher homotopy Lie algebras (see \cite{Stasheff1997a} for further references). Classical string field theory can be seen as an implementation of the framework presented here, although I've actually worked the other way, abstracting what is non-particular to strings. The various algebraic structures, on the other hand, can be considered as (itself quite abstract, but understood) semantical domains for the syntax presented here. In fact, there is an enormous amount of mathematics that might be relevant. We don't understand higher spin gauge fields well enough to target the appropriate mathematics with precision yet, although it seems possible to semantically map to higher homotopy Lie algebras. Still we need a concrete, physics like, implementation. Thus, I envisage a four-tier structure: syntactic formulation of the theory $\rightarrow$ semantical map onto a known mathematical structure $\rightarrow$ concrete physics-like implementation $\rightarrow$ computational implementation.

\section{Notes on the free field theory}\label{sec: Free}
The free field theory of higher spin gauge fields is a simple and quite beautiful generalization of the lower spin gauge theories for spin 1 and spin 2. There are lots of variations, developed and proposed during the last three decades \cite{Penrose1965,Fronsdal1978,Shirafuji1983,Curtright1985,deWitFreedman,Vasiliev1980a,DamourDeser1987}. Indeed, the literature on higher spin fields is enormous, and I will not attempt to review it. However, it seems to me that certain approaches stand out as particularly simple and, perhaps for that reason, suited as a basis for interactions. One is the original formulation \cite{Dirac1936a,FierzPauli,Chang,SinghHagen} in terms of field equations and Lagrangians for symmetric tensor fields, perfected by Fronsdal \cite{Fronsdal1978}. Another is the formulation of Freedman and deWit \cite{deWitFreedman} in terms of generalized Christoffel symbols (see also \cite{Curtright1979}). A third is the BRST approach developed by Siegel and Zwiebach \cite{SiegelZwiebach1987}, S. Ouvry and J. Stern \cite{OuvryStern1987a} and independently by myself {\cite{AKHB1987a}. This approach was inspired by and adapted from the, at that time, very active work on string field theory. The reader is refered to the review \cite{CBThorn1989} for a list of references. The BRST approach has been rediscovered \cite{Labastida1989,PashnevTsulaia1998a,Bonelli2003a,FranciaSagnotti2003a}\footnote{There might be more recent papers on the subject of which I'm unaware. I apologize for any omissions. Some of these rediscoveries have been made in the context of the tensionless string, also referred to in the 1986 works.} several times during the years since it was first wrote down in 1986. Then there is the light-front formulation \cite{BBB1983a,BBB1983b,BBL1987}, which, in a way, is the simplest formulation of all, were it not for the intricate mixing of gauge symmetry with Poincar{\'e} symmetry which make higher order interactions untractable. Then we have the approach of M. Vasiliev \cite{Vasiliev1980a} which has lead to a dramatic progress on interactions in an AdS background geometry. See \cite{Vasiliev2004Review} and \cite{Vasiliev1999Review} for recent reviews and further references, as well as a view on the higher spin problem that is complementary to the one given in the present paper.

For completeness, the twistor approach should be noted \cite{Penrose1965,Penrose1977}.

A detailed discussion of the BRST formulation of the free field theory, suitable for our present purposes, can be found in \cite{AKHB1988}. Here, I will just review the notation that will be needed when discussing the BRST implementation.

Consider a phase space spanned by bosonic variables $(x_\mu,p_\mu)$ and $(\alpha_\mu,\alpha^{\dagger}_\mu)$ and ghost variables $(c^+,b_+)$, $(c^-,b_-)$ and $(c^0,b_0)$ with commutation relations

\begin{eqnarray}
[x_{\mu},p_{\nu}]=i\eta_{\mu\nu},\quad[\alpha_{\mu},\alpha_{\nu}^{\dagger}]=\eta_{\mu\nu}\label{eq: BosonicCommRels}\\
\{c^+,b_+\}=\{c^-,b_-\}=\{c^0,b_0\}=1\label{eq: GhostCommRels}.
\end{eqnarray}

The ghosts have the following properties under hermitean conjugation
\begin{equation}\label{eq: GhostConjugation}
(c^-)^\dagger=c^+,\quad(b_-)^\dagger=b_+,\quad(c^0)^\dagger=c^0,\quad(b_0)^\dagger=b_0.
\end{equation}

The vacuum is degenerate
\begin{eqnarray}\label{eq: DegenerateVacuum}
\alpha_\mu|+\rangle=\alpha_\mu|-\rangle=0 \\
\langle+|-\rangle=\langle-|+\rangle=1 \\
\langle+|+\rangle=\langle-|-\rangle=0
\end{eqnarray}
with the properties
\begin{eqnarray}\label{eq: VacuumProp}
b_0|+\rangle=0,\quad b_0|-\rangle=|+\rangle, \\
c^0|-\rangle=0,\quad c^0|+\rangle=|-\rangle, \\
c^-|+\rangle=c^-|-\rangle=0, \\
b_+|+\rangle=b_+|-\rangle=0.
\end{eqnarray}

The ghost variables are Grassmann odd, while the bosonic are Grassmann even. The equations $b_0|-\rangle=|+\rangle$ and $c^0|+\rangle=|-\rangle$ relating the vacua, then implies that either one of the two vacua must be odd. I will choose $|-\rangle$ Grassmann even and $|+\rangle$ Grassmann odd. A peculiar consequence is that $\langle+|-\rangle$ becomes odd.

Ghost numbers, ${\rm gh()}$, are assigned according to
\begin{equation}\label{eq: GhostNumbers}
\cases{x_\mu,p_\mu,\alpha_\mu,\alpha^\dagger_\mu\quad\quad 0\cr
c^0,c^+,c^-\quad\quad\quad\quad 1\cr
b_0,b_+,b_-\quad\quad\quad -1\cr
|+\rangle\quad\quad\quad\quad\quad-1/2\cr
|-\rangle\quad\quad\quad\quad\quad\quad1/2\cr}.
\end{equation}

The higher spin fields are collected into the ket $|\Phi\rangle$ with expansion
\begin{equation}\label{eq: FieldGhostExpansnon}
|\Phi\rangle=\Phi(p)|+\rangle+F(p)c^+b_{-}|+\rangle+H(p)b_-)|-\rangle,
\end{equation}
where $\Phi(p)$ contains the symmetric higher spin gauge fields, and $F(p)$ and $H(p)$ are certain auxiliary fields. These fields are further expanded in terms of the oscillators
\begin{eqnarray}
\Phi=\Phi_0+i\Phi^\mu\alpha^\dagger_\mu+\Phi^{\mu\nu}\alpha^\dagger_\mu\alpha^\dagger_\nu+\ldots,\label{eq: FieldOscillatorExpansion1} \\
F=F_0+iF^\mu\alpha^\dagger_\mu+F^{\mu\nu}\alpha^\dagger_\mu\alpha^\dagger_\nu+\ldots, \label{eq: FieldOscillatorExpansion2}\\
H=H_0+iH^\mu\alpha^\dagger_\mu+H^{\mu\nu}\alpha^\dagger_\mu\alpha^\dagger_\nu+\ldots.\label{eq: FieldOscillatorExpansion3}
\end{eqnarray}

The gauge parameters are expanded as
\begin{equation}\label{eq: GaugeParameter}
|\Xi\rangle=(\xi_0-i\xi^\mu\alpha^\dagger_\mu+\xi^{\mu\nu}\alpha^\dagger_\mu\alpha^\dagger_\nu+\ldots)b_-|+\rangle.
\end{equation}

Note that the field $|\Phi\rangle$ is odd while the $|\Xi\rangle$ is even. The Grassmann properties are carried by the vacua. When abstracting the free theory, the vacua will be dropped, and the Grassmann properties of $\Phi$ and $\Xi$ will be interchanged. In principle, they can be defined either way in the abstract theory.

The BRST operator $Q$ is expressed in terms of the generators
\begin{equation}\label{eq: Generators}
G_0={1\over 2}p^2,\quad G_-=\alpha\cdot p,\quad G_+=\alpha^{\dagger}\cdot p,
\end{equation}
spanning the simple algebra
\begin{equation}\label{eq: Algebra}
[G_-,G_+]=2G_0,
\end{equation}
with all other commutators zero.

In terms of these generators, the BRST operator reads\footnote{Signs are choosen so that the component field actions works out as in reference \cite{WeinbergQFT1}.}
\begin{equation}\label{eq: BRSToperator}
Q=c^0G_0-c^+G_+ -c^-G_- -2c^+c^-b_0.
\end{equation}

The action
\begin{equation}\label{eq: FreeAction}
A=\langle\Phi|Q|\Phi\rangle,
\end{equation}

is invariant under the gauge transformations,
\begin{equation}\label{eq: FreeGaugeTransformation}
\delta_\Xi|\Phi\rangle=Q|\Xi\rangle,
\end{equation}
as is the field equation
\begin{equation}\label{eq: FreeFieldEquation}
Q|\Phi\rangle=0.
\end{equation}

There is one, somewhat puzzling aspect, of the this theory. When expanding out the equations (\ref{eq: FreeAction}), (\ref{eq: FreeGaugeTransformation}) and (\ref{eq: FreeFieldEquation}), everything works out nicely for the component fields, except the fact that the theory contains auxiliary fields, all of which cannot be solved for without introducing a further constraint. This constraint is applied to both the field and the gauge parameter
\begin{equation}\label{eq: TConstraint}
T|\Phi\rangle=0,\quad T|\Xi\rangle=0,
\end{equation}
where $T$ is the operator
\begin{equation}\label{eq: TOperator}
T={1\over 2}\alpha\cdot\alpha+b_+c^-.
\end{equation}

When expanded, the constraint equations works out to the double tracelessness constraint for component fields of spin $s\geq 4$ and the tracelessness constraint for the corresponding component gauge parameters. The free field theory is still gauge invariant without imposing these constraints, but the constraints are needed in order to get the correct number of physical degrees of freedom. These questions are discussed in \cite{AKHB1988}.

So the bottom line of the BRST treatment of the free field theory is that the action can be written as $\langle\Phi|Q|\Phi\rangle$ with $Q$ a nilpotent kinetic operator and $|\Phi\rangle$ a certain expansion over internal, ghost degrees of freedom. Below, when discussing an abstract approach to the theory, I will not use this notation, instead inventing a new syntax. The reason is to keep bra and ket vectors, oscillators and commutator brackets et cetera, where they belong, namely in the implementation.

It is neither particularly difficult, nor very interesting at the present stage of investigation to write down more complicated free field theories than the one reviewed here. For our present purposes, it suffices to consider this free theory, which is presumable the simplest one, as an example to abstract.

\subsubsection*{A note on units}
Working in mass units, i.e. with dimensionality ${\rm d}(p)=1$, all configuration space fields have ${\rm d}(\Phi(x))={D-2\over 2}$ where $D$ is the spacetime dimension, while momentum space fields have ${\rm d}(\Phi(p))=-{D+2\over 2}$. All in all we get dimensionalities, ${\rm d}()$, according to
\begin{equation}\label{eq: Dimensionalities}
\cases{p,c^+,c^-\quad\quad\quad\; 1 \cr
\alpha,\alpha^\dagger,c^0,b^0\quad\quad 0\cr
|+\rangle,|-\rangle\quad\quad\quad\;\; 0 \cr
x,b^+,b^-\quad\quad -1.\cr} 
\end{equation}
Consequently, ${\rm d}(|\Phi\rangle)=-{D+2\over 2}$ and ${\rm d}( Q)=2$.

\section{Notes on self-interactions}\label{sec: Inter}
There is a large amount of work on interactions for higher spin gauge fields. I will certainly not try to review the full body of knowledge on the matter. Instead, I will outline a few salient features that I think are significant with respect to the self-interaction problem. I will do this based on a brief overview of the subject, well aware of the prejudices this might entail.\footnote{In particular, I will leave out all reference to work on coupling point particles to background higher spin gauge fields, not because the topic is uninteresting, but it lies somewhat outside the main thrust of the present paper. The same comment goes for all occurences of massless higher spin fields in string theory and its relatives.}

As far as I know, the question of introducing self-interactions\footnote{Electromagnetic and gravitational interactions will be briefly discussed at the end of this section.} for higher spin massless fields was first published in the paper \cite{FangFronsdal1979}, reviewing the so named 'Gupta' program. This program belongs to the attempts to quantize gravity in terms of an interacting spin 2 field \cite{Gupta1952,Gupta1954}. This approach can be entered via two paths, either by linearizing Einstein gravity, or by starting with a free spin 2 field and attempting to introduce self-consistent  interactions in an iterative way. This latter approach soon acquired an immediacy on its own, more or less independent of the quantization problem. Indeed, the question of self coupling a free spin 2 field into a non-linear theory can be studied as a problem in classical field theory. Some of the often cited early work on this subject are \cite{Kraichnan1955,Wyss1965,Thirring1961,Feynman1962a}. 

That general relativity can be derived from requiring a consistent self-interacting theory of spin two fields, was shown by S. Deser in \cite{Deser1970} (see also \cite{BoulwareDeserKay1979}. In this paper, it is shown that the Einstein and Yang-Mills theories can both be derived from the requirement of self-interaction in just one iterative step. The resulting theories are cubic in the first order form, i.e. where pairs of independent fields $(g^{\mu\nu},\Gamma^\alpha_{\mu\nu})$ and $(A_\mu,F_{\mu\nu})$ are used respectively. In this approach, the further non-linearities of the theories are hidden in the choice of first order field variables, since upon making connection to the standard formulation, the field $\Gamma^\alpha_{\mu\nu}$ must be solved in terms $g^{\mu\nu}$, and $F_{\mu\nu}$ in terms of $A_\mu$ in the well known way. This derivation uses only the Abelian local gauge invariances of the free theories, and the full non-Abelian local gauge invariances are a result rather than an input.

But in the course of all this work on Yang-Mills theory and gravity, it became apparent that gauge invariance was a crucial concept. By iteratively adding non-linear terms to the free spin 2 action and Abelian gauge transformations, it could be proved \cite{FangFronsdal1979} that a consistent, gauge invariant theory of self-interacting spin two fields can be constructed that is equivalent to Einstein gravity. These derivations relied on starting from Minkowski spacetime, and a free spin 2 field $h_{\mu\nu}$ propagating in this flat background. That this condition of Minkowski background could be lifted was shown in \cite{Deser1987}. 

It was only natural then to try and extend the program to spin 3 fields and higher. The general idea behind this approach is to take a free field theory and its Abelian gauge symmetry and then {\it deform} it into a non-linear theory. The problem was from the outset \cite{FangFronsdal1979} put in a deformation theoretic context \cite{Gerstenhaber}. Most authors express the hope that the so constructed non-linear theory, if it exists, will turn out to be unique. A first requirement for this method to succeed is that a general enough ansatz for the non-linear theory can be written down. This is a non-trivial problem in general. As soon as we pass spin 2, the problem explodes in a potential complexity of fields, multiplets, background manifolds, dimensions, symmetry groups, et cetera. 

Returning to the historical path, at roughly the same time, another approach to gravity appeared, more closely modeled on Yang-Mills theory. The Yang-Mills interaction of spin 1 fields was introduced by 'gauging' of a global symmetry algebra like $SU(N)$ in the well-known way \cite{YangMills1954}. A natural question was whether gravity could likewise be obtained by gauging an appropriate global spacetime symmetry. The candidate, global spacetime symmetries are the Lorentz or Poincar{\'e} symmetries. It turns out that gravity can indeed be obtained by gauging the Poincar{\'e} group using techniques of vierbeins and spin connections \cite{Utiyama,Kibble,MacDowellMansouri1977,MansouriChang,Grensing}. The 'gauging' approach is different from the 'deformation' approach in that the gauge algebra is fixed, but it is promoted from being global to becoming local. In deformation theory, the algebra is already local, but it is promoted from being Abelian to becoming non-Abelian. The question then is whether the two approaches yield the same result. In the case of spin 1 the answer is definitely positive \cite{OgievetskyPolubarinov1962,OgievetskyPolubarinov1963}. In the case of spin 2, there are conceptual problems involved, at least if one is strongly prejudiced towards a geometrical view, but barring this, the results agree. 

The subsequent work on the higher spin problem followed these two paths; deforming the free theory Abelian gauge group, or gauging a global (but non-Abelian) symmetry algebra. The impressive work of M. Vasiliev falls in the second category. The far less advanced BRST program falls in the first category, as does the light-front approach and the approach of Berends, Burgers and van Dam referred to below. 

Note also that these methods have been extensively used for obtaining various supergravity theories (for reviews, see \cite{Nieuwenhuizen,DuffNilssonPope}).

As regards higher spin gauge field interactions, the first positive result was the light-front construction of cubic interaction terms for arbitrary spin \cite{BBB1983a}. 

When going from spin 1 and spin 2 in the light-front approach to higher spin, nothing strange happened. Quite to the contrary, the generalization seemed very natural. In four dimensions, where each and every integer spin gauge field has just two physical helicity degrees of freedom, parameterized by a natural number $\lambda$, the cubic interaction term can be written
\begin{equation}\label{eq: CubicInteractionLightFront}
g\int d^4x\sum_{n=0}^g(-1)^n{\lambda\choose n}(\partial^+)^\lambda\phi\Big[{\partial\over\partial^+}\Big]^{(\lambda-n)}{\overline\phi}\Big[{\partial\over\partial^+}\Big]^n{\overline\phi}+\mbox{complex conjugate},
\end{equation} 
where the two components of the complex field $(\phi,\overline \phi)$ corresponds two the helicities $\lambda$ and $-\lambda$, and where $(\partial,\overline\partial)$ are complex transverse partial derivatives. The interaction is essentially a binominal expansion. In the case of odd $\lambda$, the fields entering the interaction term carry an index $\in\{a,b,c\}$ contracted into an anti-symmetric symbol $f_{abc}$ reminiscent of the situation for spin 1. How this generalizes to higher orders in the interaction is not known, as the quartic interaction term resisted attempts at construction at the time.

From the cubic interaction term, we can read of the following information
\begin{itemize}
\item There are $\lambda$ transverse derivatives.
\item The coupling constant $g$ has mass dimension $1-\lambda$.
\item The odd spin fields carry an anti-symmetrized index.
\end{itemize}

The covariant spin 3 vertex constructed in \cite{BerendsBurgersvanDam1984} is consistent with these general properties. Furthermore, fixing the light-front gauge in the covariant cubic interaction for spin 3 yields precisely the light-front cubic interaction term for spin 3 \cite{IBunpubl1}.
 
In a way, the light-front result is a bit odd. If it turns out, as claimed by M. Vasiliev, that higher spin interactions requires an anti-deSitter background, then why does the nonlinear Minkowski Poincar\' e algebra allow this term? Is it just a coincidence, and the theory breaks down at the quartic level? We do not know. Note, though, that there are other cubic interaction terms on the light-front, involving fields of different helicities \cite{BBL1987}. Whether any of these terms describes higher spin interaction with gravity to lowest order in the spin 2 field, is at least to my knowledge, not known. The same situation occurs in covariant approach of Berends, Burgers and van Dam \cite{BerendsBurgersvanDam1984}. 

These latter authors found that upon commuting two spin 3 gauge transformations, the commutator did not close on spin 3, but produced terms that could be interpreted as gauge transformations of for fields of spin $>3$. This is clear hint that once one goes beyond spin 2, an infinite tower of higher spin fields will be needed (see also \cite{Fronsdal1979conf} and \cite{AKHB1985}). Berends, Burgers and van Dam \cite{BerendsBurgersvanDam1985,Burgers1985thesis} furthermore made an extensive analysis of the higher spin problem that is still highly relevant. Their analysis is within the 'deformation' approach and is based on the 'original' formulation of the free field theory mentioned in section \ref{sec: Free}. 

\subsubsection*{Higher spin fields in electromagnetic and gravitational backgrounds}
Self interactions was not the first type of interaction discussed for higher spin fields. Rather, it was electromagnetic and  gravitational interactions. To begin with, massive higher spin fields, i.e. matter, in particular spin $3/2$, was studied. Later the discussion included massless higher spin gauge fields. There is a huge literature on this subject, and I will just point out a list of original references (hopefully not to incomplete) as well as some recent papers that might be helpful to the reader wishing to pursue this topic. 

To make a long story short, minimally coupling of higher spin gauge fields to gravity violates the higher spin gauge invariances. In the special case of supergravity, non-minimal terms can be added that saves the theory. But in general, higher spin fields coupled to gravity suffers from a consistency problems that cannot be alleviated by non-minimal couplings. Similar problems arise in attempts to couple higher spin fields to electromagnetism. There thus seems to be no consistent way of introducing higher spin fields into a pre-existing spin 1 and 2 system. The problems was first noted already by Fierz and Pauli \cite{FierzPauli}, and there exist a long series of papers discussing these problems \cite{Buchdahl1958,Buchdahl1962,JohnsonSudarshan,VeloZwanziger,AragoneDeser1971,AragoneDeser1980a,AragoneDeser1979,AragoneDeser1980b,BerendsHoltenWitNieuwenhuizen,BarthChristensen,AragoneLaRoche,Curtright1980}. A modern reference is \cite{DeserWaldron2002}.

In view of these discouraging results it is reasonable to ask if it is at all useful to pursue investigations into higher spin gauge field interactions. My own point of view is based on the following three observations, and I'm aware of the fact that this is a weak spot; (i) negative (so called 'no-go' results) have been circumvented before, (ii) as soon as spin 2 is passed, all spins must be included, and presumably all be treated on common ground, and it is not clear what happens then, (iii) all negative results derives within a spacetime setting, this might be misguided if higher spins plays any fundamental role at all.

The reader should note that the present paper does not purport to solve these problems, but rather proposes a way to work around them, by setting up a framework with as few as possible restrictions.

\section{Abstracting dynamics}\label{sec: AbsDynamics}
Physics concerns itself with the dynamics of physical systems. A physical system is a part of the universe with a well defined interface towards the rest of the universe which becomes the environment. One of the standard paradigms of dynamics is to describe the system in terms of an action. The action in its turns depends on a set of dynamical variables. Often the action possess symmetries, i.e. parameter dependent variations of the variables that leave the action essentially invariant. Equations of motion are obtained by varying the action with respect to the dynamical variables. The scheme is well known to every physicist, and clearly, it can be formalized. Here I will choose a moderate level of formalization, sufficient as a backdrop to the formalization we will need for higher spin gauge fields.

Thus, abstract dynamics can be described as follows.

Let the description of the system be in terms of a set of variables $\{\phi_i\}$ where the index $i$ runs over an index set ${\cal I}$. Dynamics is governed by the action $A$. The action is a function of the variables $A(\{\phi_i\})$. The equations of motion follows from varying the action with respect to the variables. Formally we have
\begin{equation}\label{eq: AbsVaryingAction}
\forall j\in{\cal I}:\big(\delta_{\phi_j}A(\{\phi_i\})=0\rightarrow \exists W_j:W_j(\{\phi_i\})=0\big).
\end{equation}

Here, W denotes the equations of motion
\begin{equation}\label{eq: AbsEqMotionGen}
\forall j\in {\cal I}:W_j(\{\phi_i\})=0.
\end{equation}

Defining the variation $\delta_{\phi_j}$ requires some care, but I will rely on the standard application of this operation.

Invariance of the action under symmetry transformations
\begin{equation}\label{eq: AbsSymmetryTrans}
\delta_\xi\phi_i=f(\{\phi_n\},\xi),
\end{equation}

is the demand that the variation of the action evaluates to zero
\begin{equation}\label{eq: AbsInvarianceAction}
\delta_\xi A=0.
\end{equation}

We also demand that the transformation close and form an algebra, possibly modulo the field equations,
\begin{eqnarray}\label{eq: AbsGaugeAlgebra}
\lbrack\delta_{\xi_1},\delta_{\xi_2}\rbrack\phi=f(\{\delta_{\xi_1}\phi_n\},\xi_2)-f(\{\delta_{\xi_2}\phi_n\},\xi_1) \nonumber\\
=\delta_{\xi(\{\phi_n\},\xi_1,\xi_2)}\phi\;({\rm mod}\;W).
\end{eqnarray}

The algebra can be field dependent, as signaled by the field dependent gauge parameter $\xi(\{\phi_n\},\xi_1,\xi_2)$ in the commutator of two gauge transformations.

\section{Abstracting the free field theory}\label{sec: AbsFree}
All higher spin gauge fields, as well as auxiliary fields, are packaged into one master field $\Phi$. Let such a field be called a HS field. The explicit representation is left to the implementation. However, we need a way to extract the component fields. Let us write this formally as
\begin{equation}\label{eq: FormalGetField1}
{\bf get}(\Phi,s)\rightarrow\phi_s,
\end{equation}
where we can think of ${\bf get}$ as either an operator acting on $\Phi$ or as a function call in the case that the theory is implemented computationally. ${\bf get}$ applied to the field equation for $\Phi$ should yield the component field equations. In practice, in order to make contact with conventional field theory, the component fields will be ordinary symmetric tensors, so that
\begin{equation}\label{eq:FormalGetField1}
\phi_s(p)=\phi_{\mu_1,\mu_2,\ldots\mu_s}(p).
\end{equation}

Here, $p$ denotes a momentum space coordinate and the indices $\mu$ are spacetime indices. It should be kept in mind, though, that there might be situations where we want to hide the spacetime representation, or where we want to extract an entirely different representation. There is also the possibility that there is no spacetime representation.\footnote{A deep reason for the severe problems in constructing interactions might be that spacetime is not the proper arena for higher spins. Assuming that the very concept of spin can be given a reasonable definition independent of a background spacetime geometry, such a representation is currently under investigation.}

Furthermore, we need a way to distinguish different fields. Here we will build in one piece of classical field theory. Fields depend on variables, in general spacetime coordinates and possibly extra variables. All these will be package into one indexed symbol $\sigma_i$, which as already noted, need not be related to spacetime at all. Thus we will write HS fields as $\Phi(\sigma_i)$, sometimes abbreviated to $\Phi_i$ for convenience.

Implicit in the above discussion on the HS fields is that they belong to some set $\cal H$. Eventually $\cal H$ might be a Hilbert space, but we need not presuppose that as yet. We write $\Phi::{\cal H}$. One can think of this equation as stating the {\it type} of $\Phi$. 

Furthermore, to the extent to which we need to be able to multiply fields by numbers and add them, we may assume that the set ${\cal H}$ of HS fields is a vector space. The scalars of the vector space can be complex numbers of even or odd Grassmann parity. The fields might carry a Grassmann party $\varrho(\Phi)\in\{0,1\}$, so that
\begin{equation}\label{eq: GrassmannField}
\Phi_1\Phi_2=(-)^{\varrho(\Phi_1)\varrho(\Phi_2)}\Phi_2\Phi_1.
\end{equation}

The product involved here is just a direct product $\otimes$. This 'equation' could also be regarded as a purely textual ordering of symbols. It will be used subsequently when a proper field product is defined.

A minimum requirement in order to write down a free field theory action is that we can write a real bilinear form containing some kind of kinetic operator $K$. Contemplating this, it becomes clear that we need to enforce the structure of an inner product ${\bf in}(\cdot\,,\cdot)$ on the vector space.
\begin{equation}\label{eq: BilinearForm}
{\bf in}(\cdot\,,\cdot)::{\cal H}^2\rightarrow{\rm C}.
\end{equation}

Finally, we need a structure of linear operators acting on the fields. Let $K$ be one such operator
\begin{equation}\label{eq: KineticOperator}
K::{\cal H}\rightarrow{\cal H},
\end{equation}

then the free field theory action is written
\begin{equation}\label{eq: GenFreeAction}
A(\Phi)={\bf in}(\Phi ,K\Phi).
\end{equation}

It is clearly interesting to analyze what is the weakest possible structure that needs to be introduced. The above assumptions are more based on intuition than on a systematic study. As already noted in the introduction, a pure syntactic formulation should presumably be pursued, but it is not useful at the present exploratory stage of investigation.

In this context, it can be discussed how general the equation (\ref{eq: GenFreeAction}) for the free action really is. It is clearly an abstraction of the $\langle\Phi|Q|\Phi\rangle$ action in the BRST approach. However, that the free action should be a bilinear in the abstract field $\Phi$ is hard to dispute. Furthermore, the field equations ought to involve some operators, differential (or momentum) in a spacetime description, so it is hard to escape some (linear) operator $K$ acting on $\Phi$ and in some way 'extracting' a concrete kinetic operator acting on the component fields. But the reader has to judge for herself/himself.

\section{Abstracting interactions}\label{sec: AbsInteractions}
In any field theory, interactions between different fields enter the field equations with non-linear terms, corresponding to non-quadratic contributions to the action. To accommodate this in our scheme, we need a way to form products between fields. It is immediately clear that ordinary naive school products are insufficient in the general case. That only works in scalar polynomial theories like the $\phi^3$-model. All other field theories, electrodynamics, gravity, string field theory, et cetera, requires more elaborate schemes. For example, the Yang-Mills three field interaction term reads 
\begin{equation}\label{eq: YM3FieldintX}
gf_{abc}A^a_\mu(x) A^b(x)\cdot\partial A(x)^{c\mu}.
\end{equation}

Superficially, the three fields seem to enter the interaction term in an unsymmetric way. But if the term is transformed to momentum space, it can be written as
\begin{equation}\label{eq: YM3FieldIntP}
gf_{abc}A^a(p_1)\cdot A^b(p_2)(p_1-p_2)\cdot A^{c}(p_3)+\mbox{cyclic permutations}.
\end{equation}
The configuration space interaction term is local in spacetime, whereas in momentum space, the fields entering the interaction term carry their own momenta. This is the form of interaction that we want to abstract, since then each field carry a unique label encoded in $\sigma_i$.

A product of $n$ HS fields is a multilinear map {\bf pr}\,:: ${\cal H}^{\otimes n}\rightarrow\cal H$
\begin{equation}\label{eq: FieldProduct}
\Phi(\sigma_{n+1})={\bf pr}\big(\Phi(\sigma_1),\Phi(\sigma_2),\ldots,\Phi(\sigma_{n})\big).
\end{equation}
A priori, this product has no symmetries, an issue to which we will return below. A shorthand notation is useful when the field arguments are not needed
\begin{equation}\label{eq: ShortHandProduct}
{\bf pr}(\Phi^n)\equiv{\bf pr}\big(\Phi(\sigma_1),\Phi(\sigma_2),\ldots,\Phi(\sigma_{n})\big)\equiv{\bf pr}(\Phi_1,\ldots,\Phi_n).
\end{equation}

We will also need expressions like ${\bf pr}(\Phi^k,\Psi^l)$, which are naturally expanded as need be
\begin{equation}\label{eq: GroupOfFields}
{\bf pr}(\Phi^k,\Psi^l)={\bf pr}(\Phi_1,\ldots,\Phi_k,\Psi_1,\ldots,\Psi_l)
\end{equation}

Multilinearity entails
\begin{eqnarray}\label{eq: Multilinearity}
{\bf pr}(\Phi_1,\ldots,a_n\Phi_n+b_n\Psi_n,\ldots,\Phi_m)\nonumber\\
=a_n(-)^{\iota(a_n,n)}{\bf pr}(\Phi_1,\ldots,\Phi_n,\ldots,\Phi_m)\nonumber\\
+b_n(-)^{\iota(b_n,n)}{\bf pr}(\Phi_1,\ldots,\Psi_n,\ldots,\Phi_m).
\end{eqnarray}
where 
\begin{equation}\label{eq: Grassmann1}
\iota(c_n,n)={\varrho(a_n)(\varrho(\Phi_1)+\ldots +\varrho(\Phi_{n-1})}
\end{equation}
Upon reordering adjacent fields in the product, there might be a sign flip in the case where they anticommute.
\begin{eqnarray}\label{eq: SignFlip}
{\bf pr}(\Phi_1,\ldots,\Phi_n,\Psi_{n+1},\ldots,\Phi_m)\nonumber\\
=(-)^{\varrho(\Phi_n)\varrho(\Psi_{n+1})}{\bf pr}(\Phi_1,\ldots,\Psi_{n+1},\Phi_n,\ldots,\Phi_m).
\end{eqnarray}

The HS fields themselves can be chosen as Grassmann even, as seen from the free field theory (where the odd parity derives from the vacuum). But we need this generality, taking grading into account, since the gauge parameters are odd, and there will occur odd operators like $Q$. This also derives from the BRST free theory. The product itself is assumed to carry no intrinsic Grassmann parity. 

Now, flipping the order of adjacent fields in the product, all permutations of the fields can be reached. In the case of Grassmann even fields, the product is therefore independent of the particular order in which the fields are written, and the product is strictly commutative. This might sound confusing, since we certainly do not expect an Abelian theory. However, non-commutativity enters when one considers the nested products that appear when studying gauge invariance and commutators of gauge transformations. Furthermore, associativity is not an issue at this stage. Here we are considering primitive $n$-ary products, and associativity only enters upon discussing, for example, expressing a product of three factors in terms of successive products of two factors. 

It should be clear that in order for this product to serve as a basis for introducing interactions, it must have further properties. This is precisely the subject of our study. The bare minimum of such properties will be derived from the requirement of gauge invariance of the action.

The low indices $n=0$ and $n=1$ merit simplified notation. Thus we define
\begin{equation}\label{eq: LowIndex0}
{\bf pr}(\Phi^0)\equiv{\bf pr}()=0,
\end{equation}
where $\Phi^0$ is defined to be a void argument, and 
\begin{equation}\label{eq: LowIndex1}
{\bf pr}(\Phi)=K\Phi.
\end{equation}

Since for $n=1$, {\bf pr} is of type ${\cal H}\rightarrow\cal H$, it makes sense to define it as linear transformation. If any of the fields is identically zero for all values of the labels $\sigma_i$, then the product is zero.

We now have the abstract tools for writing interactions. By taking the product between $n-1$ HS fields and then the inner product with an $n$-th field, a candidate for an $n$-field interaction term can be written
\begin{equation}\label{eq: InteractionTerm}
{\bf in}\big(\Phi_{n+1},{\bf pr}(\Phi_1,\Phi_2,\ldots,\Phi_n)\big).
\end{equation}

It makes sense to introduce a special notation for this expression
\begin{equation}\label{eq: DefVX}
{\bf vx}(\Phi_1,\Phi_2,\ldots,\Phi_n)={\bf in}\big(\Phi_n,{\bf pr}(\Phi_1,\Phi_2,\ldots,\Phi_{n-1})\big)
\end{equation}
so that ${\bf vx}$ is a multilinear map ${\bf vx}::{\cal H}^{\otimes n}\rightarrow{\rm C}$.

Just as the product (\ref{eq: FieldProduct}), ${\bf vx}$ has no a priori symmetries. However, if this is to useful in an interaction term, it must at least be cyclic symmetric in the field (compare to the Yang-Mills three field interaction term above (\ref{eq: YM3FieldIntP})). This can be fixed by explicitly summing over all permutations of the fields. Alternatively, one could just sum over all cyclic permutations. It turns out, though, that summing over all permutations leads to simpler formulas. Computationally, it is inefficient to sum over all permutations. However, in actual implementations, the permutation symmetries will be explicit, and the combinatorial sums collapse into at most ${\cal O}(n^2)$ terms.}

To that end, let $\pi[0..n]$ denote the set of all permutations of the list $[0..n]$ of natural numbers between 0 and $n$. Then $\sum_{\pi[n]}$ will denote a sum over permutations.

This finishes the setting up of the basic syntax of the theory. In the following sections, we will perform calculations within this syntax. That might seem a bit strange, since we have not defined any rules of calculation or rewrite rules. However, close scrutiny of the manipulations that follow show that we only need to do substitutions and rearrangements of sums. This is a weak form of equational reasoning that we certainly want to do in any formalism, but strictly speaking, rewrite rules, belong to semantics.

\section{The action and gauge invariance}\label{sec: Action}
Since we are working within the 'deformation' tradition it makes sense to write an ansatz for the action as a formal power series in the polymorphic map ${\bf vx}$. The action then reads
\begin{equation}\label{eq: FullAction1}
A(\Phi)=\sum_{i=2}^\infty{g^{i-2}\over{i!}}\sum_{\pi[i]}{\bf vx}(\Phi_1,\Phi_2,\ldots,\Phi_i).
\end{equation}
To clean up notation, introduce a new summation symbol $\sum_{\pi(i=m)}^\infty\equiv\sum_{i=m}^\infty\sum_{\pi[i]}$.

The action can written, highlighting the kinetic term explicitly and expanding ${\bf vx}$, in the form 
\begin{eqnarray}\label{eq: FullAction2}
A(\Phi)={\bf in}(\Phi,K\Phi)+\sum_{\pi(i=3)}^\infty{g^{i-2}\over{i!}}{\bf in}(\Phi_i,{\bf pr}(\Phi_1,\Phi_2,\ldots,\Phi_{i-1})\nonumber\\
={\bf in}(\Phi,K\Phi)+\sum_{\pi(i=3)}^\infty{g^{i-2}\over{i!}}{\bf in}(\Phi,{\bf pr}(\Phi^{i-1})),
\end{eqnarray}
where 
\begin{equation}\label{eq: KineticTermExplicit}
{\bf in}(\Phi,K\Phi)={1\over 2}({\bf in}(\Phi_1,K_2\Phi_2)+{\bf in}(\Phi_2,K_1\Phi_1)).
\end{equation}

The gauge transformation will also be written as a formal power series. At this stage we have a choice either to introduce a new abstract product ${\bf pr_g}$ (renaming the previously introduced product ${\bf pr_a}$), or use the same product as the one used for the action. This is precisely one of the points where one is confronted with a dilemma as to the generality of the ansatz. I will be conservative here and use the same product. The gauge transformation then reads
\begin{equation}\label{eq: GaugeTrans}
\delta_{\Xi}\Phi=\sum_{\pi(i=0)}^\infty{g^i\over{i!}}{\bf pr}(\Phi^i,\Xi)=K\Xi+\sum_{\pi(i=1)}^\infty{g^i\over{i!}}{\bf pr}(\Phi^i,\Xi).
\end{equation}

Gauge invariance of the action to all orders of interaction amounts to
\begin{equation}\label{eq: DeltaActionZero}
\delta_{\Xi}A(\Phi)=0.
\end{equation}

By demanding this to be true, we can derive the requisite demands on the maps ${\bf in}$, ${\bf pr}$ and ${\bf vx}$. We have to go through this calculation meticulously as we have to record all steps where required properties of the maps has to recorded. To that aim, apply $\delta_{\Xi}$ to the action (\ref{eq: FullAction1})
\begin{equation}\label{eq: VaryAction1}
\delta_{\Xi}A(\Phi)=\sum_{i=2}^\infty{g^{i-2}\over{i!}}\delta_{\Xi}\sum_{\pi[i]}{\bf vx}(\Phi^i).
\end{equation}

We immediately run into the problem of how to perform this operation. However, by analyzing how the operation of varying the action is normally done in standard field theories, we see that this can be done by just textually substituting $\delta_{\Xi}\Phi$ for all the occurences of $\Phi$ one at a turn so to speak. Indeed
\begin{eqnarray}\label{eq: VaryProduct1}
\delta_{\Xi}{\bf vx}(\Phi_1,\Phi_2,\ldots,\Phi_i)={\bf vx}(\delta_{\Xi}\Phi_1,\Phi_2,\ldots,\Phi_i)\nonumber\\
+{\bf vx}(\Phi_1,\delta_{\Xi}\Phi_2,\ldots,\Phi_i)+\ldots +{\bf vx}(\Phi_1,\Phi_2,\ldots,\delta_{\Xi}\Phi_i),
\end{eqnarray}

so that we get
\begin{equation}\label{eq: Varyproduct2}
\delta_{\Xi}\sum_{\pi[n]}{\bf vx}(\Phi^n)=n\sum_{\pi[n]}{\bf vx}(\delta_{\Xi}\Phi,\Phi^{n-1}).
\end{equation}

Thus continuing the calculation (\ref{eq: VaryAction1}), we get
\begin{equation}\label{eq: VaryAction2}
\delta_{\Xi}A(\Phi)=\sum_{\pi(i=2)}^\infty{g^{i-2}\over{(i-1)!}}{\bf vx}(\delta_{\Xi}\Phi,\Phi^{i-1}).
\end{equation}

Then , upon substituting (\ref{eq: GaugeTrans}) for $\delta_{\Xi}\Phi$ and shifting the $i$-sum, $i\rightarrow i-1$
\begin{equation}\label{eq: VaryAction3}
\sum_{\pi(i=1)}^\infty{g^{i-1}\over{i!}}\sum_{\pi(j=0)}^\infty{g^j\over{j!}}{\bf vx}\Big({\bf pr}(\Phi^j,\Xi),\Phi^i\Big).
\end{equation}

Here, we can use the definition (\ref{eq: DefVX}) of ${\bf vx}$ 
\begin{equation}\label{eq: VaryAction4}
\sum_{\pi(i=1)}^\infty\sum_{\pi(j=0)}^\infty{g^{i+j-1}\over{i!\,j!}}{\bf in}\Big({\bf pr}(\Phi^j,\Xi),{\bf pr}(\Phi^i)\Big).
\end{equation}

No harm is done by extending the $i$-sum to start at $i=0$. Assuming that it is allowed to rearrange the double sum, it can be written as
\begin{equation}\label{eq: VaryAction5}
\sum_{i=0}^\infty\sum_{{\pi(k=0)\atop \pi(l=0)}}^{k+l=i}{g^{i-1}\over{k!\,l!}}{\bf in}\Big({\bf pr}(\Phi^k,\Xi),{\bf pr}(\Phi^l)\Big),
\end{equation}
i.e. we are summing order by order in the total power of the field, corresponding to how an order by order checking of invariance would be performed. Thus, for any fixed $i=n$, study
\begin{eqnarray}\label{eq: VaryAction6}
\sum_{{\pi(k=0)\atop \pi(l=0)}}^{k+l=n}{g^{n-1}\over{k!\,l!}}{\bf in}\Big({\bf pr}(\Phi^k,\Xi),{\bf pr}(\Phi^l)\Big)\nonumber\\
=\sum_{{\pi(k=0)\atop \pi(l=0)}}^{k+l=n}{g^{n-1}\over{k!\,l!}}{\bf in}\Big({\bf pr}(\Phi^l),{\bf pr}(\Phi^k,\Xi)\Big).
\end{eqnarray}

Again using the definition (\ref{eq: DefVX}) of ${\bf vx}$
\begin{eqnarray}\label{eq: VaryAction7}
\sum_{{\pi(k=0)\atop \pi(l=0)}}^{k+l=n}{g^{n-1}\over{k!\,l!}}{\bf vx}\Big({\bf pr}(\Phi^l),\Phi^k,\Xi\Big)\nonumber\\
=\sum_{{\pi(k=0)\atop \pi(l=0)}}^{k+l=n_0}{g^{n-1}\over{k!\,l!}}{\bf vx}\Big(\Xi,\Phi^k,{\bf pr}(\Phi^l)\Big)\nonumber\\
=\sum_{{\pi(k=0)\atop \pi(l=0)}}^{k+l=n_0}{g^{n-1}\over{k!\,l!}}{\bf in}\Big(\Xi,{\bf pr}(\Phi^k,{\bf pr}(\Phi^l))\Big).
\end{eqnarray}

Normally, invariance should not depend on the gauge parameter $\Xi$, so in order for this sum to vanish, we must require for all $n\in {\rm N}$
\begin{equation}\label{eq: ProductIdentity}
\sum_{{\pi(k=0)\atop \pi(l=0)}}^{k+l=n}{1\over{k!\,l!}}{\bf pr}(\Phi^k,{\bf pr}(\Phi^l))=0.
\end{equation}

This is a non-trivial demand on the map ${\bf pr}$. The other demands can be considered as part of the syntax, but this one involves the semantics of the theory. I will refer to this requirement as the {\it product identity}. 

\subsubsection*{Low level special cases of the product identity}
The first four levels, i.e. values of $n$, are of immediate importance.

When $n=0$ the equation trivializes to 
\begin{equation}\label{eq: N=0}
{\bf pr}({\bf pr}())={\bf pr}(0)=0.
\end{equation}

The case $n=1$ becomes
\begin{equation}\label{eq: N=1}
{\bf pr}(\Phi,{\bf pr}())+{\bf pr}({\bf pr}(\Phi))=KK\Phi=0.
\end{equation}

This equation expresses gauge invariance for the free theory. 

When $n=2$, taking permutations into account and noting that ${\bf pr}(\Phi_1,\Phi_2)$ is actually symmetric, we get
\begin{equation}\label{eq: N=2}
{\bf pr}({\bf pr}(\Phi_1,\Phi_2))+{\bf pr}(\Phi_1,{\bf pr}(\Phi_2))+{\bf pr}(\Phi_2,{\bf pr}(\Phi_1))=0,
\end{equation}
or
\begin{equation}\label{eq: N=2K}
K{\bf pr}(\Phi_1,\Phi_2)+{\bf pr}(\Phi_1,K\Phi_2)+{\bf pr}(K\Phi_1,\Phi_2)=0,
\end{equation}

This equation expresses gauge invariance of the cubic interaction term. Granting that we already know how to implement the $n=1$ equation in terms of an appropriate field and a nilpotent kinetic operator, the $n=2$ equation is the first non-trivial equation to implement. It involves the two-product ${\bf pr}(\,\cdot,\cdot)$ which so far is undefined. This is the product that was partially studied in \cite{AKHB1988}. 

The next level, $n=3$ involves the quartic interaction term
\begin{eqnarray}\label{eq: N=3K}
K{\bf pr}(\Phi_1,\Phi_2,\Phi_3)\nonumber\\
+{\bf pr}(K\Phi_1,\Phi_2,\Phi_3)+{\bf pr}(\Phi_1,K\Phi_2,\Phi_3)+{\bf pr}(\Phi_1,\Phi_2,K\Phi_3)\nonumber\\
+{\bf pr}(\Phi_1,{\bf pr}(\Phi_2,\Phi_3))+{\bf pr}(\Phi_2,{\bf pr}(\Phi_3,\Phi_1))+{\bf pr}(\Phi_3,{\bf pr}(\Phi_1,\Phi_2))=0.
\end{eqnarray}

This equation expresses gauge invariance up to the quartic level. In order to solve it, the full two-product ${\bf pr}(\,\cdot,\cdot)$ must have been obtained first. Clearly, it can then be seen as a 'differential' equation for the three-product ${\bf pr}(\,\cdot,\cdot,\cdot)$ with $K$ acting as differential operator.

It follows that a necessary condition for the interaction to be cubic is that the two-product satisfies a Jacobi identity. In that case, the three-product ${\bf pr}(\,\cdot,\cdot,\cdot)$ is  zero, and the first four terms in (\ref{eq: N=3K}) vanishes. Thus 
\[
\mbox{Cubic interaction}\quad\rightarrow\quad\mbox{Two-product satisfies Jacobi identity}.
\]

\subsubsection*{The field equations}

Varying the action (\ref{eq: FullAction1}) with respect to the field $\Phi$ yields the field equation
\begin{eqnarray}\label{FieldEquation1}
\delta_{\Phi}A(\Phi)=\sum_{\pi(i=2)}^\infty{g^{i-2}\over{(i-1)!}}{\bf vx}(\delta_{\Phi},\Phi^{i-1})\nonumber\\
={\bf in}\Big(\delta_{\Phi},\sum_{\pi(i=1)}^\infty{g^{i-1}\over{i!}}{\bf pr}(\Phi^i)\Big).
\end{eqnarray}

According to the abstract dynamics, we get the field equation
\begin{equation}\label{FieldEquation2}
W(\Phi)=\sum_{\pi(i=1)}^\infty{g^{i-1}\over{i!}}{\bf pr}(\Phi^i)=0,
\end{equation}
thus expressing the basic intuition that the product captures the interactions.

\section{The gauge algebra}
In order to examine the gauge algebra, we do the standard calculation
\begin{eqnarray}\label{eq: GaugeCommutator}
\lbrack\delta_{\Xi_1},\delta_{\Xi_2}\rbrack\Phi=\delta_{\Xi_1}\delta_{\Xi_2}\Phi-(1\leftrightarrow2)\nonumber\\
=\delta_{\Xi_1}\Big(\sum_{\pi(i=0)}^\infty{g^i\over{i!}}{\bf pr}(\Phi^i,\Xi_2)\Big)-(1\leftrightarrow2)\nonumber\\
=\sum_{\pi(i=1)}^\infty{g^i\over{(i-1)!}}{\bf pr}(\delta_{\Xi_1}\Phi,\Phi^{i-1},\Xi_2)-(1\leftrightarrow2)\nonumber\\
=\sum_{\pi(i=0)}^\infty\sum_{\pi(j=0)}^\infty{g^{i+j+1}\over{i!j!}}{\bf pr}\big({\bf pr}(\Phi^j,\Xi_1),\Phi^i,\Xi_2)\big)-(1\leftrightarrow2)\nonumber\\
=\sum_{i=0}^\infty\sum_{{\pi(k=0)\atop \pi(l=0)}}^{k+l=i}{g^{k+l+1}\over{k!\,l!}}{\bf pr}\big(\Phi^k,\Xi_1,{\bf pr}(\Phi^l,\Xi_2)\big)-(1\leftrightarrow2).
\end{eqnarray}

This is as far as we can get without invoking semantics for the product. Experience with field theory, shows that the commutator of two gauge transformations should close on a new, possibly field dependent gauge transformation, and possibly modulo the field equations.

The form of equation (\ref{eq: GaugeCommutator}) suggests considering 
\begin{equation}\label{eq: }
{\bf pr}\big(\Phi^k,{\bf pr}(\Xi_1,\Xi_2,\Phi^l)\big).
\end{equation}

If we could establish the following identity
\begin{eqnarray}\label{eq: GaugeIdentity}
\sum_{{\pi(k=0)\atop \pi(l=0)}}^{k+l=i}{g^{k+l+1}\over{k!\,l!}}\Big\lbrace{\bf pr}\big(\Phi^k,{\bf pr}(\Xi_1,\Xi_2,\Phi^l)\big)+{\bf pr}\big(\Phi^k,\Xi_1,\Xi_2,{\bf pr}(\Phi^l)\big)\nonumber\\
+{\bf pr}\big(\Phi^k,\Xi_1,{\bf pr}(\Xi_2,\Phi^l)\big)-{\bf pr}\big(\Phi^k,\Xi_2,{\bf pr}(\Xi_1,\Phi^l)\big)\Big\rbrace=0,
\end{eqnarray}
then the calculation (\ref{eq: GaugeCommutator}) could be continued with
\begin{eqnarray}\label{eq: GaugeCommutatorContinued}
-\sum_{n=0}^\infty\sum_{{\pi(k=0)\atop \pi(l=0)}}^{k+l=n}{g^{k+l+1}\over{k!\,l!}}\Big({\bf pr}\big(\Phi^k,{\bf pr}(\Xi_2,\Xi_1,\Phi^l)\big)+{\bf pr}\big(\Phi^k,\Xi_2,\Xi_1,{\bf pr}(\Phi^l)\Big)\nonumber\\
=-\sum_{\pi(k=0)}^\infty{g^k\over k!}{\bf pr}\big(\Phi^k,\sum_{\pi(l=0)}^\infty{g^{l+1}\over l!}{\bf pr}(\Xi_2,\Xi_1,\Phi^l)\big)\nonumber\\
-\sum_{\pi(k=0)}^\infty{g^k\over k!}{\bf pr}\big(\Phi^k,\Xi_2,\Xi_1,\sum_{\pi(l=0)}^\infty{g^{l+1}\over l!}{\bf pr}(\Phi^l)\big).
\end{eqnarray}

The first term can be recognized as a field dependent gauge transformation and the second as being proportional to the field equations, thus
\begin{equation}\label{eq: }
\lbrack\delta_{\Xi_1},\delta_{\Xi_2}\rbrack\Phi=\delta_{\Xi(\Phi,\Xi_1,\Xi_2)}\Phi+g^{2}\sum_{\pi(k=0)}^\infty{g^{k}\over k!}{\bf pr}(\Phi^k,\Xi_1,\Xi_2,W(\Phi)),
\end{equation}
where the new gauge parameter is
\begin{equation}\label{eq: GaugeParameter}
\Xi(\Phi,\Xi_1,\Xi_2)=\sum_{\pi(l=0)}^\infty{g^{l+1}\over l!}{\bf pr}(\Xi_1,\Xi_2,\Phi^l).
\end{equation}

We can now record the required properties of the product. Apart from being able to make substitutions, add and compare equal terms, i.e. use standard equational reasoning, the product identity (\ref{eq: ProductIdentity}) is the only non-trivial demand on ${\bf pr}$. The equation (\ref{eq: GaugeIdentity}) that is needed in the gauge algebra calculation, can be subsumed in a generalization of the product identity (\ref{eq: ProductIdentity}). Let us see how this can be done. 

The product identity was derived under the assumption that all the fields were Grassmann even, and that the only odd object was the gauge parameter. Therefore, the Grassmann properties could be ignored. We have to generalize (\ref{eq: ProductIdentity}) to include the case of fields with even and odd parities. To that end, let $\{\Gamma_i\}$ denote a set of $n$ fields with $\varrho(\Gamma_i)\in\{0,1\}$. 

The sum in the product identity runs over all permutations just for convenience. The different terms are cyclic permutations of the split of the string of fields $\Phi_1\cdots\Phi_n$ into two strings with $k$ and $l$ fields respectively, and the factor ${1/k!l!}$ is cancelled against the number of equal permutations in each split. Therefore (\ref{eq: ProductIdentity}) can be written as
\begin{equation}\label{eq: ProductIdentity2}
\sum_{{k=0,l=0}\atop \rm{cycl. perm.}}^{k+l=n}{\bf pr}(\Phi^k,{\bf pr}(\Phi^l))=0.
\end{equation}

Another way to express this is to consider the index set $\{1,\cdots,n\}$ as split into the two sets $\{i_1,\cdots, i_k\}$ and $\{j_1,\cdots, j_l\}$. 

Denote the split $\{\{i_1,\cdots, i_k\},\{j_1,\cdots, j_l\}\}$ by $\chi(k,l)$. The sum then runs over all different such splits
\begin{equation}\label{eq: ProductIdentity3}
\sum_{{k=0,l=0}\atop \chi(k,l)}^{k+l=n}{\bf pr}(\Phi_{i_1}\cdots\Phi_{i_k},{\bf pr}(\Phi_{j_1}\cdots\Phi_{j_l}))=0.
\end{equation}

The order of the indices does not matter when all the fields are Grassmann even. When arbitrary parities are involved, we need a convention as to the order. Instead of sets $\{i_1,\cdots i_k\}$ and $\{j_1,\cdots j_l\}$ we use ordered lists $\lbrack i_1..i_k\rbrack$ and $\lbrack j_1..j_l\rbrack$. This pair of ordered lists is denoted by $\pi(k,l)$. It is a particular permutation of the index set into two lists with $i_1<\cdots<i_k$ and $j_1<\cdots<j_l$.

If the fields carry arbitrary Grassmann parity, the weakest generalization is to record a sign picked up when the order of the fields $\Gamma_{i_1}\cdots\Gamma_{i_k}\cdots\Gamma_{j_1}\cdots\Gamma_{j_l}$ are reordered 'lexicographically' into $\Gamma_1\Gamma_2\cdots\Gamma_n$. Denote this sign by $\epsilon(\pi(k,l))$.

We have to sum over all such splittings, keeping track of the signs
\begin{equation}\label{eq: ProductIdentityGeneral}
\sum_{{k=0,l=0}\atop \pi(k,l)}^{k+l=n}\epsilon(\pi(k,l)){\bf pr}(\Gamma_{i_1}\cdots\Gamma_{i_k},{\bf pr}(\Gamma_{j_1}\cdots\Gamma_{j_l}))=0.
\end{equation}

We can then return to the equation (\ref{eq: GaugeIdentity}) that governs the closure of the gauge algebra. Consider first the last expression of the commutator calculation (\ref{eq: GaugeCommutator}). Expanding the permutations we have 
\begin{equation}\label{eq: GaugeCommutator2}
\sum_{{k=0,l=0}\atop \rm{cycl. perm.}}^{k+l=n}{\bf pr}\big(\Phi^k,\Xi_1,{\bf pr}(\Phi^l,\Xi_2)\big)-(1\leftrightarrow2).
\end{equation}

Consider applying the identity (\ref{eq: ProductIdentityGeneral}) to the string of fields $\Gamma^n=\Phi^{k'}\Xi_1\Xi_2\Phi^{l'}$. Of the terms in the sum, there are terms where both $\Xi_1$ and $\Xi_2$ are in the first list, both $\Xi_1$ and $\Xi_2$ in the second list, and terms where $\Xi_1$ ($\Xi_2$) is in the first and $\Xi_2$ ($\Xi_1$) in the second. Writing this out explicitly yields
\begin{eqnarray}\label{eq: IdExplicit}
\sum_{{k=0,l=0}\atop \chi(k,l)}^{k+l=n}\Big\lbrace{\bf pr}(\Phi^k,{\bf pr}(\Xi_1,\Xi_2,\Phi^l))+{\bf pr}(\Phi^k,\Xi_1,\Xi_2,{\bf pr}(\Phi^l))\nonumber\\
+{\bf pr}(\Phi^k,\Xi_1,{\bf pr}(\Xi_2,\Phi^l))-{\bf pr}(\Phi^k,\Xi_2,{\bf pr}(\Xi_1,\Phi^l))\Big\rbrace=0.
\end{eqnarray}
Thus, the identity (\ref{eq: GaugeIdentity}) we used in the gauge algebra calculation follows from the generalized product identity (\ref{eq: ProductIdentityGeneral}) taking Grassmann parities into account. 

Therefore, given a particular type of free gauge fields $\Phi$, gauge parameter $\Xi$, a kinetic operator $K$ and the inner product ${\bf in}$, construction of the map ${\bf pr}$ satisfying (\ref{eq: ProductIdentityGeneral}) is the only non-trivial task. 

In this way, the vaguely defined problem of introducing interactions for higher spin gauge fields, has been focused on implementing the product map ${\bf pr}$ satisfying the identities in equation (\ref{eq: ProductIdentityGeneral}). Since very little has been assumed as to the particulars of such an implementation, we are quite free to explore various implementational schemes. These vary from finding pre-existing mathematical domains to setting up concrete data structures within which computerized explorations can be performed. In the last two paragraphs, I will outline examples of these two 'extreme' approaches.

\section{Strongly homotopy Lie algebras}
The product identities we have found are similar to the defining identities for strongly homotopy Lie algebras (sh-Lie algebras or ${\rm L}(\infty)$ algebras) \cite{LadaStasheff1993a,LadaMarkl1995a}. This opens the possibility to map the syntax set up here onto such an algebra. The problem is that we do not know which particular algebra to choose. 

There are a few variants of the basic definitions of strongly homotopy Lie algebras in the literature, but the following, mildly technical, is sufficient for our purpose to bring out the similarity to the product identity.

\subsubsection*{Definition}
Consider a ${\bf Z}_2$ graded vector space $V=V_0\oplus V_1$ over some number field, and denote the elements by $x$. The grading is given by $\varrho$ with $\varrho(x)=0$ if $x\in V_0$ and $\varrho(x)=1$ if $x\in V_1$. $V$ is supposed to carry a sequence of $n$-linear products denoted by brackets. The graded  $n$-linearity is expressed by
\begin{equation}\label{eq: LinearProducts1}
[x_1,\ldots,x_n,x_{n+1},\dots,x_m]=(-)^{\varrho(x_n)\varrho(x_{n+1})}[x_1,\ldots,x_{n+1},x_n,\dots,x_m]
\end{equation}
\begin{eqnarray}\label{eq: LinearProducts2}
\lbrack x_1,\ldots,a_nx_n+b_nx'_n\,\ldots,x_m\rbrack\nonumber\\
=a_n(-)^{\iota(a_n,n)}\lbrack x_1,\ldots,x_n,\ldots,x_m\rbrack+b_n(-)^{\iota(b_n,n)}\lbrack x_1,\ldots,x'_n,\ldots,x_m\rbrack
\end{eqnarray}
where $\iota(a_n,n)={\varrho(a_n)(\varrho(x_1)+\ldots +\varrho(x_{n-1})}$.

The defining identities for the algebra are, for all $n\in{\rm N}$
\begin{equation}\label{eq: DefiningEquations}
\sum_{{k=0}\atop{l=0}}^{k+l=n}\sum_{\pi(k,l)}\epsilon(\pi(k,l))\lbrack\lbrack x_{\pi(1)},\ldots,x_{\pi(k)}],x_{\pi(k+1)},\ldots,x_{\pi(k+l)}\rbrack=0.
\end{equation}
where $\pi(k,l)$ stands for (k,l)-unshuffles. A (k,l)-unshuffle is a permutation $\pi$ of the indices $1,2,\cdots,k+l$ such that $\pi(1)<\ldots<\pi(k)$ and $\pi(k+1)<\ldots<\pi(k+l)$. $\epsilon(\pi(k,l))$ is the sign picked up during the unshuffle as the points $x_i$ with indices $0\leq i\leq k$ are taken through the points $x_j$ with indices $k+1\leq j\leq l$. This is just the normal procedure in "superalgebras".

The low index, $n=0$ and $n=1$ brackets are treated separately, thus
\begin{eqnarray}
\lbrack\,\cdot\,\rbrack=0 \\
\lbrack x\rbrack=\partial x, 
\end{eqnarray}
with $\partial$ a derivation.

This is a definition. Given that such algebras do exist, it is clear that they offer a possible semantic target for abstract higher spin gauge fields. It is obvious that the image of a field $\Phi_n$ is a point $x_n$, and that the products ${\bf pr}(\cdot)$ maps into the brackets  $\lbrack\,\cdot\,\rbrack$. A technical detail is that in order for the mapping to be complete, the sh-Lie algebra must be supplied with an inner product.

The details of setting up this mapping would require some care, but it should be essentially straightforward. The problem lies elsewhere, we still do not know which particular algebra to map to. Had we known the correct concrete algebra, then we would have had a solution to the higher spin problem.

A way out of this dilemma is to, as a first step, map the corresponding categories instead. By formalizing the syntax given here, a category of interacting fields, say ${\bf IField}$ is set up. The same is done for strongly homotopy Lie algebras, denoted by ${\bf shLie}$. The interpretation map $[|\cdot|]$ is then a functor from the interacting fields to the sh-Lie algebras.
\[
[|\cdot|] :: {\bf IField}\rightarrow {\bf shLie}.
\]

Clearly, there is much work to done here and many technical details to work out. It should be noted that category theory can in fact be used in denotational (mathematical) semantics for programming languages \cite{Pierce}. Given that a programming language is just an example of a formal language, and that abstract field theory also can be formulated as a formal language, it should be clear that this point of view is feasible. Whether it helps in the quest to obtain a concrete physics-like implementation remains to be seen. At least, we are able to put interacting higher spin gauge fields in a context where models can be systematically searched for. 

From a physics point of view it is clear that the abstract view given here must be supplemented by physical insight into the problem. That is, what are higher spin gauge fields? What kind of physics do they describe? What is the proper context to set them in? Lacking that understanding, I will in the next section set up a framework for a concrete implementation within which the problem can at least be pursued by brute force computerized calculation.

\section{Vertex implementation}
With the free field theory implemented using the BRST technique briefly reviewed in section \ref{sec: Free}, it is natural to try to implement the interacting field products in terms of vertex operators. This is how string field theory is done. Taking the open string as an example, the three string vertex is a product of a bosonic vertex and a ghost vertex. The bosonic vertex is 
\[
|V_3\rangle=\exp\Big({1\over 2}\sum_{r,s=1}^3\sum_{n,m=0}^\infty\alpha_{-n}^{\mu,r}N_{nm}^{rs}\alpha_{-m}^{\mu,s}\Big)|-\rangle_3,
\]
in terms of bosonic string oscillators $\alpha_{-n}^{\mu,r}$ and the Neumann function matrices $N_{nm}^{rs}$. There has been attempts to use this form of three vertex for higher spin gauge fields, but that fails since such a vertex do not reproduce the spin 1 Yang-Mills cubic interaction terms \cite{KohOuvry}. It is known from the light-front formulation cubic interaction terms \cite{BBL1987} that the vertex must at least contain terms of the generic form $\alpha^{\dagger}\alpha^{\dagger}\alpha^{\dagger}p$ (indices suppressed), i.e. with three oscillators and one momentum label. Such a covariant vertex was partially determined in \cite{AKHB1988} and it correctly reproduces the Yang-Mills cubic interaction term. Further progress was halted by a lack of an effective way of calculating higher order terms in the three vertex. In this section, higher spin vertices will be discussed from the point of view of the abstract approach presented here.

The object is to define the field products in terms of vertex operators. An $n$-vertex operator is an object that takes $n-1$ fields, labeled by $\{\sigma_i\}_1^{n-1}$ and which outputs a new field, labeled by $\sigma_n$. Each field $\Phi(\sigma_i)$ is represented as a ket vector $|\Phi(\sigma_i)\rangle$ in the oscillator and ghost Fock space corresponding to the label $\sigma_i$ as in equations (\ref{eq: FieldGhostExpansnon}), (\ref{eq: FieldOscillatorExpansion1}), (\ref{eq: FieldOscillatorExpansion2}) and (\ref{eq: FieldOscillatorExpansion3}). The product is defined by
\begin{equation}\label{eq: VertexFieldProduct}
{\bf pr}(\Phi(\sigma_1),\ldots,\Phi(\sigma_{n-1}))\equiv\langle\Phi(\sigma_1)|\cdots\langle\Phi(\sigma_{n-1})||V(\sigma_1,\ldots,\sigma_{n-1},\sigma_n)\rangle.
\end{equation}

The notation makes it explicit that the product evaluates to a ket field $|\Phi(\sigma_n)\rangle$, or
\begin{equation}\label{eq: EvaluateVertex}
\langle\Phi(\sigma_1)|\cdots\langle\Phi(\sigma_{n-1})||V(\sigma_1,\ldots,\sigma_{n-1},\sigma_n)\rangle\rightarrow|\Phi(\sigma_n)\rangle.
\end{equation}

Note that the fields $|\Phi\rangle$ are now Grassmann odd due to the vacuum $|+\rangle$, whereas the gauge parameters are Grassmann even. As will be seen below, the vertex can be built from Grassmann even objects, so that it has full permutational symmetry in all the $n$ field labels.

Likewise the map ${\bf vx}$ is represented by
\begin{equation}\label{eq: VertexInteraction}
{\bf vx}(\Phi(\sigma_1),\ldots,\Phi(\sigma_n))\equiv\langle\Phi(\sigma_1)|\cdots\langle\Phi(\sigma_n)||V(\sigma_1,\ldots,\sigma_n)\rangle,
\end{equation}
which evaluates to a combination of component fields and momentum labels. In the abstract action, ${\bf vx}$ is summed over all permutations of the fields. If we allow ourselves the trick of moving all the $\langle+|_n$ vacua to the left, we get
\begin{eqnarray}\label{eq: SumPerm1}
{1\over n!}\sum_{\pi[1..n]}{\bf vx}(\Phi(\sigma_1),\ldots,\Phi(\sigma_n))\nonumber\\
={1\over n!}\Big(\otimes_{i=1}^n\langle+|_i\Big)\Big(\sum_{\pi[1..n]}\Phi(\sigma_1)\cdots\Phi(\sigma_n)\Big)|V(\sigma_1,\ldots,\sigma_n)\rangle\nonumber\\
=\langle\Phi(\sigma_1)|\cdots\langle\Phi(\sigma_n)||V(\sigma_1,\ldots,\sigma_n)\rangle,
\end{eqnarray}
i.e. one term, in that particular order. With ${\sb {1\cdots n}}\langle+|$ as a shorthand for $\otimes_{i=1}^n\langle+|_i$, this can also be written keeping all the vacuu to the left
\begin{equation}\label{eq: SumPerm2}
{\sb {1\cdots n}}\langle+|\Phi(\sigma_1)\cdots\Phi(\sigma_n)|V(\sigma_1,\ldots,\sigma_n)\rangle,
\end{equation}
a form that is convenient for explicit calculations.

When implementing the abstract gauge transformation, some care is needed considering the permutations. The vertex implementation of the gauge transformation becomes
\begin{eqnarray}\label{eq: VertexTransformation}
\delta_\Xi|\Phi(\sigma_n)\rangle=Q(n)|\Phi(\sigma_n)\rangle\nonumber\\
+{n\over 2}\sum_{\rm{cycl.perm}\atop\lbrack\sigma_1..\sigma_{n-1}\rbrack}{\sb {1\cdots n}}\langle+|\Phi(\sigma_1)\cdots\Phi(\sigma_{n-2})\Xi(\sigma_{n-1})|V(\sigma_1,\ldots,\sigma_{n-1},\sigma_n)\rangle,
\end{eqnarray}
where the coefficient $n/2$ is an artefact of the permutations.

When evaluating expressions such as (\ref{eq: VertexFieldProduct}), (\ref{eq: VertexInteraction}) and  (\ref{eq: VertexTransformation}) the explicit oscillator and ghost representations of section \ref{sec: Free} is used. 

In order for the vertex to encode non-trivial interaction information, we introduce an $n$-ary function ${\cal F}$ of the labels $\{\sigma_i\}_1^n$.  Then we write the vertex as
\begin{equation}\label{eq: ConcreteVertex}
|V(\sigma_1,\ldots,\sigma_n)\rangle={\cal F}(\sigma_1,\ldots,\sigma_n)\int d^Dp_1|-\rangle_1\cdots\int d^Dp_n|-\rangle_n \delta^D\Big(\Sigma_{i=1}^n p_i\Big),
\end{equation}
where $\delta^D\Big(\Sigma_{i=1}^n p_i\Big)$ enforces momentum conservation.

For notational convenience, write
\begin{equation}\label{eq: VertexVacuum}
|-\rangle_{1\cdots n}=\int d^Dp_1|-\rangle_1\cdots\int d^Dp_n|-\rangle_n \delta^D\Big(\Sigma_{i=1}^n p_i\Big).
\end{equation}

The $n$-order interaction term can now be written
\begin{equation}\label{eq: VertexShorthand}
g^{n-2}\langle\Phi|^{\otimes n}{\cal F}_n|-\rangle_{1\cdots n}.
\end{equation}

With this form for the vertex, mass dimensions and ghost number counting should work out correctly. It is natural to demand ${\rm gh}({\cal F})=0$, i.e. the ghost number zero is zero, but it must carry mass dimension, as will be calculated shortly.

The ghost number count works out
\[
{\rm gh}(\langle\Phi|^{\otimes n })+{\rm gh}(|-\rangle_{1\cdots n})=n(-{1\over 2})+{n\over 2}=0.
\]

The mass dimension count yields
\begin{equation}\label{eq: MassDimCount1}
(n-2){\rm d}(g)+{\rm d}( {\cal F}_n)+n(-{D+2\over 2})+nD-D=0.
\end{equation}
There is no compelling reason to let $g$ carry non-zero dimensionality, thus we set ${\rm d}(g)=0$, so that
\begin{equation}\label{eq: MassDimCount2}
{\rm d}( {\cal F}_n)=D+n-{nD\over 2}.
\end{equation}
in four dimensions, the dimension is simply $4-n$.

\subsubsection*{Ansatz for the vertex function}
The ansatz for the vertex function ${\cal F}_n$ is based on the following clauses
\begin{itemize}
\item ${\rm gh}({\cal F}_n)=0$,
\item ${\cal F}_n$ does not contain annihilators $c^0$, $c^-$, $b_+$ or $\alpha_\mu$,
\item ${\cal F}_n$ is a spacetime scalar,
\item ${\rm d}({\cal F}_n)={1\over 2}(2D+2n-nD)$.
\end{itemize}
The first three clauses imply that ${\cal F}_n$ can be built from the following bilinears
\[
\alpha^\dagger_r\cdot\alpha^\dagger_s,\quad\alpha^\dagger_r\cdot p_s,\quad c^+_rb_{s-},\quad c^+_rb_{s0},
\]
where the indices $r,s$ label HS fields. The fourth clause requires that we introduce at least one dimensional constant $\kappa$ to balance the dimensions for the second and last bilinear. Choose ${\rm d}(\kappa)=-1$. Introduce a symbol $\eta_{rs}^a$ to denote the dimensionless bilinears according to
\begin{equation}\label{eq: VertexBilinears}
\eta_{rs}^1=\alpha^\dagger_r\cdot\alpha^\dagger_s,\quad\eta_{rs}^2=\kappa\alpha^\dagger_r\cdot p_s,\quad \eta_{rs}^3=c^+_rb_{s-},\quad \eta_{rs}^4=\kappa c^+_rb_{s0}.
\end{equation}

As already noted, the higher spin vertices cannot be built out of these bilinears alone, rather, powers of the bilinears must be considered. To that end, introduce a symbol $\Delta_{2m}^n$ where $n$ denotes the order of the vertex and $m$ denotes the homogenous power of bilinears,
\begin{equation}\label{eq: VertexDeltas}
\Delta_{2m}^n={\sp n}Y^{r_1s_1\cdots r_ms_m}_{a_1\cdots a_m}\eta_{r_1s_1}^{a_1}\cdots\eta_{r_ms_m}^{a_m},
\end{equation}
where there are implicit summations according to
\begin{itemize}
\item All $r_n$ and $s_n$, $n\in[1..m]$ are summed over the list $[1..n]$
\item All $a_n$, $n\in[1..m]$ are summed over the list $[1..4]$
\end{itemize}
and where the coefficients ${\sp n}Y^{r_1s_1\cdots r_ms_m}_{a_1\cdots a_m}$ are algebraic numbers to be determined.

Finally, ${\cal F}_n$ can be synthesized as
\begin{equation}\label{eq: }
{\cal F}_n=\sum_m^\infty\kappa^{({nD\over 2}-D-n)}\Delta_{2m}^n.
\end{equation}

In this framework, the denumerable set of functions $\{{\cal F}_n\}_{n=3}^\infty$, if they exist, encode the full interacting theory of higher spin gauge fields. This same information can therefore also be considered as encoded into the denumerable set of numbers $\{{\sp n}Y^{r_1s_1\cdots r_ms_m}_{a_1\cdots a_m}\}$. 

Summarizing, we have the action
\begin{equation}\label{eq: VertexAction}
A=\langle\Phi|Q|\Phi\rangle+\sum_{n=2}^\infty g^{n-2}\langle\Phi|^{\otimes n}{\cal F}_n|-\rangle_{1\cdots n},
\end{equation}

and the gauge transformations
\begin{equation}\label{eq: VertexTransformations}
\delta_\Xi|\Phi\rangle=Q|\Xi\rangle+\sum_{n=3}^\infty{n\over 2}g^{n-2}\sum_{\rm cycl.perm}{\sb {1\cdots n}}\langle+|\Phi^{\otimes(n-2)}\Xi\,{\cal F}_n|-\rangle_{1\cdots n}.
\end{equation}

It is clear that when formulated in this manner, the gauge invariance of the action can be checked order by order by computerized calculation. At least it should be possible to work out the quartic vertex up to and including spin 3 fields. This would make it possible to compare with the known spin 1 and spin 2 cubic and quartic interaction terms, and with the covariant cubic interaction term for spin 3 derived by Berends, Burgers and van Dam \cite{BerendsBurgersvanDam1984}. Furthermore, it is likely that any obstructions that may make the theory inconsistent should crop up beyond the cubic term. 

In order to organize such a calculation a few more points needs clarifying. These are the issues of field redefinitions, global symmetries and the tracelessness constraints. A brief discussion of these points can be found in \cite{AKHB1988}. Setting up the concrete data structures can be done in a functional language like Haskell. But the details of this belongs to a computer science journal rather then a physics journal. I therefore defer a thourough discussion to a more appropriate context.

\section{Conclusions and outlook}
It is clear that the framework constructed here is not specific to higher spin gauge fields. Higher spin gauge fields enters in the specifications of the fields $\Phi$ and the kinetic BRST operator $Q$, thus essentially in the free field theory. The rest of the abstract structure is independent of the detailed form of $\Phi$ and $Q$. Whether there is an implementation of the structure or not, depends on the form of the free theory.

In particular, the abstract structure is silent on the question of multiplet structure, global symmetries, group theory factors, et cetera. The free field theory contemplated in section \ref{sec: Free} is special in that it contains just one component field of each integer spin $s$. This is perhaps the most simple situation to envision. We don't know yet if the interacting theory can be constructed in this case. It might be that more complicated multiplet structures are needed, perhaps accompanied by supersymmetry.

There is one peculiarity about the theory outlined here. What is the role of the spin 2 gauge field that appears in the free field theory? The question is connected to question of the role of gravity and spacetime background. Presumably, the free field theory can be cast in any fixed spacetime background. The kinetic operator $Q$ is known in Minkowski space and in AdS space. 

I am reluctant at the moment to speculate on a fundamental role for higher spin gauge fields. But one line of thought seem appropriate to air in the present context. The ubiquitous role of gravity at all scales of physics is one of the standard tenets of fundamental physics. In particular, this is one of the forces behind the many attempts to quantize gravity and unify gravity alongside the spin 1 Yang-Mills forces. On the other hand, there is something glaringly macroscopic about gravity. The force is weak and long range, and really just manifests itself on macroscopic scales. There is a strand of research based on the assumption that gravity is not a fundamental force at all, but just an effective force that manifests itself above sub-microscopic scales. Furthermore, as is clear from the cited work on deriving the gravitational equations either by deforming the free field theory, or by gauging the Poincar{\'e} group, the non-linearities can be understood without building it on a geometrical interpretation. This squares well with the present day folklore that spacetime breaks down at the Planck scale. But if spacetime breaks down, then so does physical geometry. This emphasizes the intuition that arithmetic is more fundamental than geometry. Arithmetic is completely scale independent, and since the abstraction of arithmetic is algebra, it can be argued that an algebraic approach to the fundamental theory is more natural than a geometric. Should it furthermore turn out that there is a fundamentally discrete substructure to reality, I think physical geometry is out at the most minute scales. 

There are no-go theorems \cite{WeinbergS1964a,WeinbergWitten1980,WeinbergQFT1} to the effect that massless fields of spin greater than 2 cannot generate long range forces. This is also consistent with everyday experience. One, admittedly sweeping, scenario would be that in a theory containing massless fields of all integer (and perhaps half-integer) spin, all the fields with spin $s>2$ just generate extreme sub-microscopic forces, while the spin 2 field gets effectively self-coupled to generate gravity, and the spin 1 fields generate the Standard Model forces, one of them surviving as long-range electrodynamics.

Related to the issues discussed here, there is one advantage of the abstract approach to higher spin fields. It makes it very natural to reconsider, as already noted, the higher spin problem in a non-spacetime context. 

To conclude, there is at least three areas where research is needed. First, the semantic mapping into generalized Lie algebras need to be clarified. Secondly, a brute force calculation within the vertex implementation should be undertaken for 'experimental' reasons. And thirdly, physical insight into the significance of higher spin gauge fields is badly needed.


\bibliographystyle{C:/TexDocs/BibStyle/unsrt}

\end{document}